\documentclass[12pt]{article}
\begin{document}
\title{Is Cold Dark Matter Baryonic? Alternative Opinion}
\author{David E. Rosenberg  Manhattanville College Purchase NY \\
\small{E-mail: rosenbergd1@gradmail.mville.edu;  \copyright 2011}}

\maketitle

\begin{abstract}
\noindent
Minutes into the big bang, nucleosynthesis finds 96 percent of matter
nonreactive. Massive pure disk protogalaxies must be 
formed in the early universe as their galaxies are present at high redshift. 
Surrounded by hot gas they made little imprint
on the CBR. The LHC has not found the Higgs or dark matter candidates.
These phenomena can be explained by a cold baryonic shell.
\end{abstract}

\section{Introduction}
Although the hierarchal $\Lambda$CDM model is quite successful on
supragalactic scales, its predictions of galactic properties differ
markedly with observation\cite{cattaneo}. 
Also general relativity has been extrapolated
to density states where the whole universe could fit into a space
smaller than an atom. There is not one shred of evidence that the
universe started at Planck densities of $10^{93} grams/cm^3$ and
temeratures of $10^{31}$ K. The nucleosynthesis of the light
elements ${  }^2H, {  }^4He,  {  }^7Li$ took place at densities of $10^5 gm/cm^3$
and about $10^{10} { }K$. Here only 4 percent of
matter was reacting. The presumption has been that all the baryons are
hot and thus the other 96 percent is nonbaryonic. Furthermore, major 
problems have been found with hierarchal galaxy formation\cite{kormendy1}.
Massive pure disk galaxies, evidently without mergers, have been found less than
$10^9$ years after the big bang. 
Dark matter detectors have not detected any nonbaryonic matter.
Additionally, the Large Hadron Collider has searched from 114 to 600 GeV
without finding the Higgs Boson or anything that could be a candidate for dark
matter\cite{editorial}.

\section{Galaxy formation}
Galaxies come in two basic types: spirals which are disk shaped and
ellipticals which are football shaped. The accepted theory is that the
present universe grew from small inhomogeneities. These grew into larger
halo structures by attracting surrounding matter. The fate of these haloes
is determined either by radiative cooling or gravitational heating. In low
mass haloes, cooling predominates, which allows cold gas to fall into the
center and become disks and stars\cite{cattaneo}. The cooling problem is
most acute in galaxies. At the end of their lives, massive stars return
$30-40$ percent of their mass to the interstellar media. If even a small
fraction of this mass is accreted, it would result in much larger black
holes. Gravitational heating dominates once a halo mass of about $10^{12}$
solar mass is reached. Cold gas is no longer able to accrete onto galaxies.
The only way galaxies within haloes can grow at this stage is by mergers.
Pure disk galaxies form bulges after the mergers. Yet samples of galaxies
have found over half are large pure disk type, without any evidence of
mergers\cite{kormendy1}.

In galaxies with bulges, the mass of the central black hole correlates with
the mass of the bulge and also the average spread of velocities of the bulge
stars. This includes ellipticals which have bulges but no disks\cite{peebles}.
An unexplained phenomenon is why the gas that formed bulge stars settled
near the black hole. Part increased the black hole mass and part led to
explosions that blew the gas away and suppressed star formation. Many
components of galaxies besides black holes are highly correlated.
The mass distribution of spiral galaxies is evenly spread from its dark
matter outer limits to its inner baryonic areas. Dark matter played a
strong role in the disk and stars but not its black hole. In pure disk
galaxies with pseudobulges, the central black hole does not correlate with
the pseudobulge\cite{kormendy3}. Another puzzle is the reason for the inward
movement of matter to the black hole in some galaxies and pseudobulge in
others. In accepted galaxy formation theory, both galaxies with and without bulges
grew by accreting matter during the period that the massive early stars were
forming. These early stars would not have settled in disks because they could
not be slowed enough to reside in disks. Galactic bulges do contain old stars
but there is no reason these old stars avoided bulgeless galaxies.
There is no evidence that they are in diffuse stellar haloes
either\cite{peebles}. Is there evidence for their existence at all?

\section{Discussion}
Could cold baryons be the proverbial dark matter?  With entirely 
hot models of the big bang such
as inflation, no such possibility could be considered. The above galaxy formation
problems are solvable assuming
there was a cold shell that formed the super massive black holes which
captured hot and cold baryons in the early universe\cite{rosenberg} 
By the time of matter-energy decoupling, the large masses with sufficient gravity were
surrounded by hot gas which prevented imprinting on the CBR. Even small black holes,
without orbiting hot gas, would not leave much of a mark. Sachs and Wolfe considered
normal matter, not galaxies and black holes, when they wrote their classic paper
on the last scattering surface for the CBR\cite{sachs}. Other solutions are as follows:

This phenomenon explains why dark
matter is highly correlated with the disk and stars but not the black hole.
There is no need to postulate the existence of massive stars in the early universe
that formed galaxies and their black holes.
Massive pure disk galaxies can be explained at high red shift without mergers.
An entirely baryonic model explains why the circular orbital speed
from luminous matter, 
which dominates the inner regions, is so similar
to dark matter at larger radii. With many stars in the center areas, initial
conditions for dark and luminous matter no longer have to be closely adjusted 
to produce a flat rotation curve. 
Hot expanding gases from the core, captured by the same size black holes
explains why there are similiar circular speeds in 
all galaxies of a given luminosity no matter how the luminous matter is spaced. 
The overall mass to light ratio rises with decreasing surface brightness so as
to preserve the Tully-Fisher relation between total luminosity and circular
speed. The depth of the gravitational well determines the circular speed
and luminosity as noted above.
The hot and cold matter discrepancies are detectable only at
accelerations below $\sim 10^{-8}{ }cm/sec$ since they are all baryons. 
The halo parameters are related to the luminous mass distribution since all
were captured by a given size black hole. The angular momentum 
of merger and collapse models is an order of magnitude less than that observed since
hierarchal dynamics were not much involved. 
The peak phase space density of the halo varies so markedly. This 
is the case since  baryonic dark matter is not collisionless and not 
homogeneously distributed. Both shell and core baryons are included. 
The predicted circular speed at a given luminosity is high. This is caused
by the capture with the slower velocity matter gaining kinetic energy
during while falling into the gravitational well. It is not directly related to 
mass to light ratio or halo density. The number of predicted subclumps in the
halos is so much greater than observed as well as predicted clumps 
compared to satellite galaxies because of the capturing process. 
Luminosity can be related to circular velocity to the fourth power as 
explained previously\cite{rosenberg}. The big bang can do without the Higgs
but a theory of everything can not.

Is it possible to explain how the big bang started with a cold shell?
Little research has been done on the fate of matter once inside a black hole\cite{hamilton}.
Between the outer and inside horizon, shearing forces may not be strong.
Black hole mathematics is only tractable if matter is being accreted slowly. 
A possible simplification is that it must cost energy to compress a neutron superfluid or
a quark-gluon plasma after all the space is eliminated.
At a limiting mass-density $\rho \approx 2.4 x 10^{16} g/cm^3$, black holes will form
\begin{equation}
\rho = \frac{c^6}{G^3M^2}
\end{equation}
above $4.5$ solar masses, not smaller. These are the smallest black holes that
have been found\cite{farr}. This phenomena can explain the gap between 
the most massive neutron stars and the smallest black holes.
When the universe underwent a collapse cycle, a massive black hole should form. 
In its interior, instead
of a singularity, there would be a very slowly expanding area of
minimal curvature. To tunnel its way out, it would take eons of exterior time.
Once it did, the universe could start anew.

\end{document}